\documentclass[a4paper,11pt]{article}
\pdfoutput=1 
\usepackage{jcappub}

\usepackage{graphicx}
\usepackage{longtable}
\usepackage{float}
\usepackage{dcolumn}
\usepackage{bm}
\usepackage{appendix}
\usepackage{multirow}
\usepackage{color}
\usepackage[utf8]{inputenc}
\usepackage{footmisc}
\newcommand{\mytilde}{\raise.17ex\hbox{$\scriptstyle\mathtt{\sim}$}}
\newcommand{\barr}{\begin{eqnarray}}
\newcommand{\earr}{\end{eqnarray}}
\newcommand{\bea}{\begin{eqnarray*}}
\newcommand{\eea}{\end{eqnarray*}}
\newcommand{\beq}{\begin{equation}}
\newcommand{\eeq}{\end{equation}}

\newcommand{\Mpc}{\mathrm{Mpc}}

\renewcommand{\bf}{\rm}
\title{CMB constraints on $\beta$-exponential inflationary models}

\author[a]{M. A. Santos} \emailAdd{aparecida.fisica@dfte.ufrn.br}

\author[b]{M. Benetti} \emailAdd{micolbenetti@on.br}

\author[a,c]{J. S. Alcaniz}\emailAdd{alcaniz@on.br}

\author[d,e]{F. A. Brito} \emailAdd{fabrito@df.ufcg.edu.br}

\author[a,f]{R. Silva} \emailAdd{raimundosilva@dfte.ufrn.br}

\affiliation[a]{Departamento de F\'\i sica, Universidade Federal do Rio Grande do Norte, 59072-970 Natal, RN, Brasil}

\affiliation[b]{Departamento de Astronomia, Observat\'orio Nacional, 20921-400, Rio de Janeiro, RJ, Brasil}

\affiliation[c]{Physics Department, McGill University, Montreal, QC, H3A 2T8, Canada}

\affiliation[d]{ Unidade Acad\^emica de F\'isica, Universidade Federal de Campina Grande, 58109-970 Campina Grande, PB, Brasil}

\affiliation[e]{Departamento de F\'isica, Universidade Federal da Para\'iba, 58051-970 Jo\~ao Pessoa, PB, Brasil}

\affiliation[f]{Departamento de F\'\i sica, Universidade do Estado do Rio Grande do Norte, Mossor\'o, 59610-210, Brasil}

\abstract{

We analyze a class of generalized inflationary models proposed in Ref.~\cite{alcaniz}, known as $\beta$-exponential inflation. We show that this kind of potential can arise in the context of brane cosmology, where the field describing the size of the extra-dimension is interpreted as the inflaton. We discuss the observational viability of this class of model in light of the latest Cosmic Microwave Background (CMB) data from the Planck Collaboration through a Bayesian analysis, and impose tight constraints on the model parameters. We find that the CMB data alone prefer \textit{weakly} the minimal standard model ($\Lambda$CDM) over the $\beta$-exponential inflation. However, when current local measurements of the Hubble parameter, $H_0$, are considered, the $\beta$-inflation model is {\it moderately} preferred over the $\Lambda$CDM cosmology, making the study of this class of inflationary models interesting in the context of the current $H_0$ tension.
}
\begin{document}
\maketitle
\section{Introduction}
\label{Introduction}

The inflationary scenario --  the current paradigm of early universe cosmology -- offers an elegant theoretical framework which is able to explain the large size and entropy of the current universe, its spatial flatness and, most importantly, the causal origin of the primordial cosmological perturbations. In the simplest models,  inflation is driven by a single minimally-coupled scalar field $\phi$ rolling down a smooth potential $V(\phi)$, which generates a primordial scalar perturbation with nearly scale-invariant power spectrum. This framework seems to agree with the most recent cosmic microwave background observations~\cite{Ade:2015xua, Ade:2015lrj}, and thanks to the accuracy of these data, it has also been possible to test the observational viability of a wide range of inflationary models (see, e.g.,~\cite{Martin:2013nzq} and references therein). However, no compelling statistical evidence has been found for a specific inflationary model and, therefore, an important task nowadays is to examine the theoretical predictions of different classes of scenarios in the light of current observational data (we refer the reader to \cite{Guth:2013sya, Ijjas:2014nta, Linde:2014nna, Brandenberger:2015kga} for different points of view of the current observational status of inflation).

In this work we study theoretical and observational predictions of a class of potentials of the type
\begin{eqnarray}
\label{potexp}
V(\phi) & = & V_0 \exp_{1-\beta}{\left(-\lambda\frac{ \phi}{M_{Pl}}\right)} \nonumber \\ & = &V_0\left[1 + \beta {\left(-\lambda\frac{ \phi}{M_{Pl}}\right)} \right]^{1/\beta}\;,
\end{eqnarray}
named $\beta$-exponential potential, which was introduced in Ref.~\cite{alcaniz} as a generalization of the usual inflationary exponential potential~(see \cite{Abbott:1984fp,Ratra:1987rm,Ferreira:1997hj} and references therein). The $\beta$-exponential function, $\exp_{1-\beta}{({f})}$, is defined as above for positive values of the term between brackets and zero otherwise,  and satisfies the inverse identity $\exp_{1-\beta}\left[\ln_{1-\beta}({f}) \right] =  {f}$, where $\ln_{1-\beta}({f}) = (f^{\beta} - 1)/\beta$ is the $\beta$-logarithmic function~\cite{alcaniz}. Although in the limit $\beta \rightarrow 0$ all the above expressions reproduce the usual exponential and logarithm properties, this is not the case for the observational predictions of the potential (1), as discussed in Ref.~\cite{Martin:2013tda}. Actually, $\beta$-exponential potentials present a number of cosmological solutions for a large interval of values of $\beta$.

The $\beta$-exponential potential was originally proposed as a purely phenomenological model. In Sec. II, however, we show how this class of potentials can arise in the context of braneworld cosmology, where the field describing the size of the extra-dimension is interpreted as the inflaton. In Sec. III we derive the observational quantities of the $\beta$-exponential inflation and discuss some prior constraints on the parameters $\beta$ and $\lambda$. A detailed observational analysis of this class of models in light of the latest CMB data provided by the Planck Collaboration~~\cite{Ade:2015xua, Ade:2015lrj} is presented in Sec. IV along with a Bayesian model comparison with respect to the standard $\Lambda$ Cold Dark Matter Inflationary ($\Lambda$CDM) cosmology. We discuss the main results of our analysis in Sec. V and summarise our main conclusions in Sec. VI. Throughout this paper we work in units such that $M_{\rm Pl} = (8\pi G)^{-1/2} = c = \hbar = 1$.


\section{ $\beta$-potentials from brane inflation}

 In what follows we show how the $\beta$-exponential potential of Eq. (1) can appear in the context of brane cosmology, where the radion (a field describing the size of the extra-dimension) is interpreted as the inflaton.  A stabilization radion stabilization mechanism is required in order to have a static extra dimension consistent with the equations of motion. This is achieved as the radion reaches the vacuum expectation value of the radion potential \cite{Dvali:1998pa,Randall:1999ee,Goldberger:1999uk,MersiniHoughton:2000ee}.

The action of the think-branes version of such scenario can be written as
\begin{equation}\label{action}
S=\int{d^Dx\sqrt{-g}\left(-\frac14 R+\frac12\partial_M\phi_i\partial^M\phi_i-V(\phi_i)\right)},
\end{equation}
which in general is a $D$-dimensional gravitational theory coupled to $i=1,2,...,N$ scalar fields, where $M=0,1,...,D-1$. Below we shall focus on $D=5$ dimensions and $N=2$ scalars. The five-dimensional geometry is assumed to have the general form with Poincar\'e invariance along the four-dimensional worldvolume of the 3-brane embedded in a five-dimensional bulk whose fifth coordinate is $r$, i.e.,
\begin{equation}
ds_5^2=e^{2A(r)}dx_\mu dx^\mu-dr^2, \qquad \mu=0,1,2,3.
\end{equation}
In a supergravity inspired action, the equations of motion can be solved by the following set of first-order differential equations for the scalar fields $\phi_i(r)$ and warp factor $\exp{(2A(r))}$
\begin{equation}\label{first-order}
A'(r)=-\frac13W(\phi_1,\phi_2),\qquad
\phi_i'(r)=\frac12 \frac{\partial W}{\partial\phi_i},  \qquad i=1,2
\end{equation}
for the scalar potential given in terms of the superpotential $W$ in the form \cite{DeWolfe:1999cp,Csaki:2000fc}
\begin{equation}\label{potential5d}
V(\phi_1,\phi_2)=\frac18\left(\frac{\partial W}{\partial\phi_i}\right)^2-\frac13W^2,\: i=1,2\;.
\end{equation}
More specifically in \cite{Fonseca:2011ep} it was addressed the scenario with dilatonic scalar potential in terms of the superpotential
\begin{equation}\label{super_pot_5d}
W(\phi_1,\phi_2)=\widetilde{W}\exp({b_1\phi_1+b_2\phi_2})\;.
\end{equation}
The {{Bogomol'nyi-Prasad-Sommerfield (BPS) solutions}} of  the first-order differential equations (\ref{first-order}) are
\begin{equation}\label{sols}
\phi_i(r)=-\frac{b_i}{b_1^2+b_2^2}A(r)
\end{equation}
with $A(r)=\ln(1+c_1r)$, for $r>0$ and $A(r)=\ln(1-c_1r)$ for $r<0$. These are dilatonic solutions \cite{Cvetic:1996vr} which are also known as {scaling solutions}  \cite{Chemissany:2007fg,Chemissany:2011gr}.

Now by properly combining background BPS solutions under the transformation $\phi_1(r)\to \phi_1(r-L)$,  which means $A(r)\to A(|r-L|)$ to patch together the solutions for $r<L$ and $r>L$ along with a {thin brane} located at $r=L$,  and  $\phi_2(r)\to -\phi_2(r)$ into the Lagrangian
\begin{eqnarray}\label{lagrangian5d}
{\cal L}_5&=&\frac12\dot\phi_1^2+\frac12\dot\phi_2^2-\frac12(\partial_r\phi_1)^2 \\ & & -\frac12(\partial_r\phi_2)^2-\widetilde{V}\exp\left({2b_1\phi_1 +2b_2\phi_2}\right)\;,  \nonumber
\end{eqnarray}
and considering that the fields have only implicit time dependence via the radion field $L\equiv L(t)$ we find
\begin{eqnarray}\label{V-1st}
{\cal L}_5&=&\frac{b_1^2}{2(b_1^2+b_2^2)^2}\frac{(\dot{L}^2-1)|r-L|'^2 c_1^2}{(1+c_1|r-L|)^2} \nonumber \\ && -\frac{b_2^2}{2(b_1^2+b_2^2)^2}\frac{c_1^2}{(1+c_1r)^2}
\nonumber \\ && -\widetilde{V}(1+c_1|r-L|)^{-\frac{2b_1^2}{b_1^2+b_2^2}}(1+c_1r)^{\frac{2b_2^2}{b_1^2+b_2^2}}.
\end{eqnarray}
At the `thin wall limit' we can identify delta functions as follows. This regime can be easily satisfied for $c_1=\frac{1}{\lambda}\frac{b_1^2+b_2^2}{2b_2^2}$ sufficiently large, where $1/\lambda$ has dimension of energy. For the sake of simplicity, we assume $b_1=\ell b_2$, such that $\frac{b_1^2}{b_1^2+b_2^2}=\frac{\ell^2}{\ell^2+1}$ and $\frac{b_2^2}{b_1^2+b_2^2}=\frac{1}{\ell^2+1}=\frac{1}{2\lambda c_1}$. As $\ell\gg1$ the Lagrangian (\ref{V-1st}) becomes
\begin{equation}\label{lagrangian5d}
{\cal L}_5=\frac12(\dot{L}^2-1)\sigma\delta(r-L)-V_0\delta(r-L)(1+c_1r)^{\frac{1}{\lambda c_1}}
\end{equation}
where $\sigma=\frac{2c_1}{b_1^2+b_2^2}$ is the brane tension.

The four-dimensional action is then given by
\begin{eqnarray}\label{action-4d}
S_4&=&\int{d^4x}\int_{-rc}^{r_c}{dr\sqrt{-g}{\cal L}_5}
\nonumber \\ &= &\int{d^4x}\sqrt{-g_4}\left(\frac12\sigma\dot{L}^2-V_{\rm eff}(L)\right)\;,
\end{eqnarray}
where $V_{\rm eff}(L)=V_0(1+c_1L)^{\frac{1}{\lambda c_1}}+\frac12\sigma$. The induced four-dimensional metric ${g_{4}}_{\mu\nu}$ is obtained from the five-dimensional metric $g_{MN}$ as follows: ${g_{4}}_{\mu\nu}(x^\mu)$=$g_{\mu\nu}(x^\mu,r=L)$, where $L$ is the position of the brane in relation to $r=0$. Particularly in the model presented above, to ensure localization of four-dimensional gravity, $L\ll r_c$, where $r_c$ is the crossover scale \cite{Fonseca:2011ep} --- for further details see \cite{Gregory:2000jc,Dvali:2000hr}.

\begin{figure*}[t]
\centering{\includegraphics[scale=0.47]{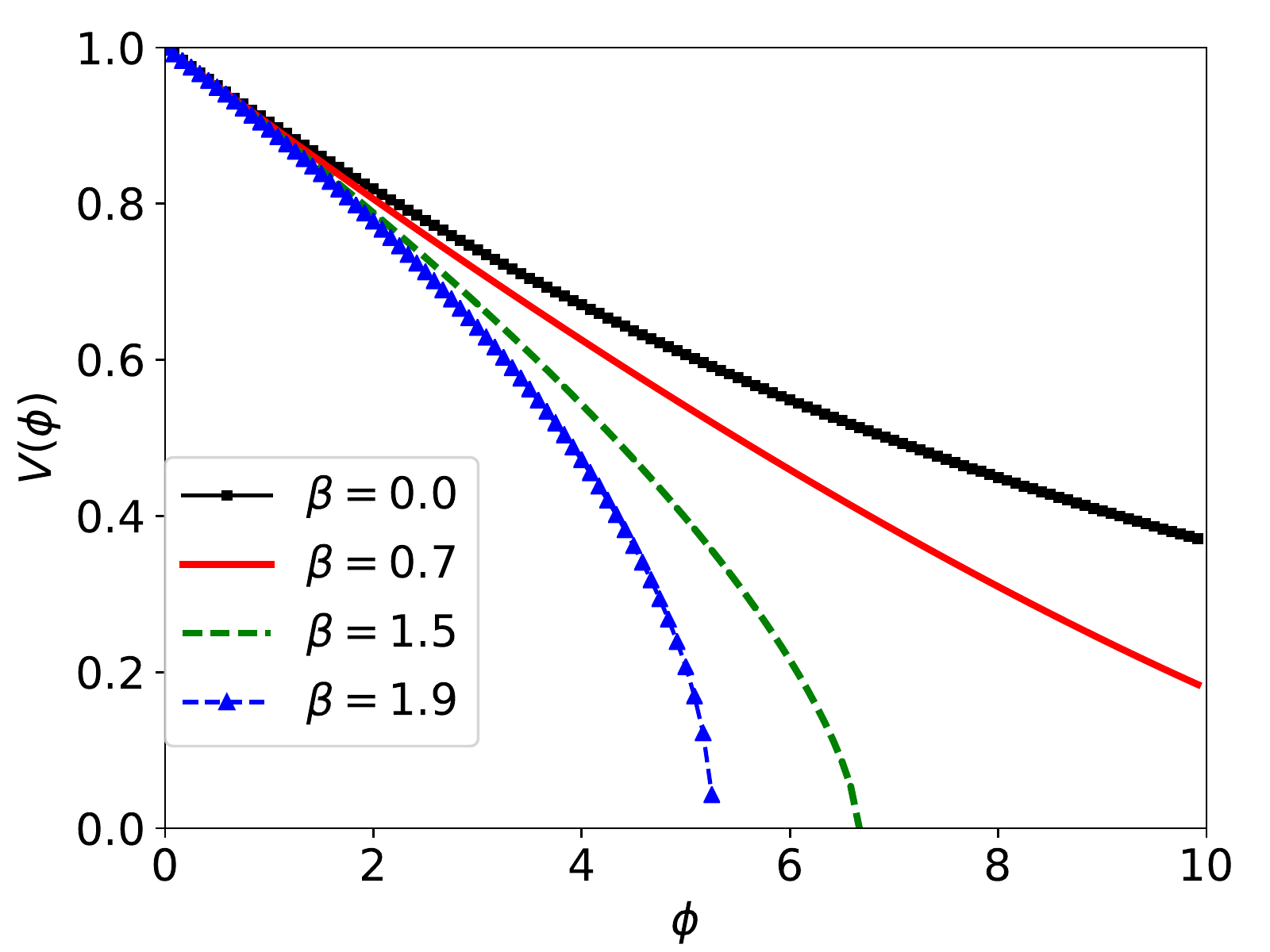}}
\centering{\includegraphics[scale=0.47]{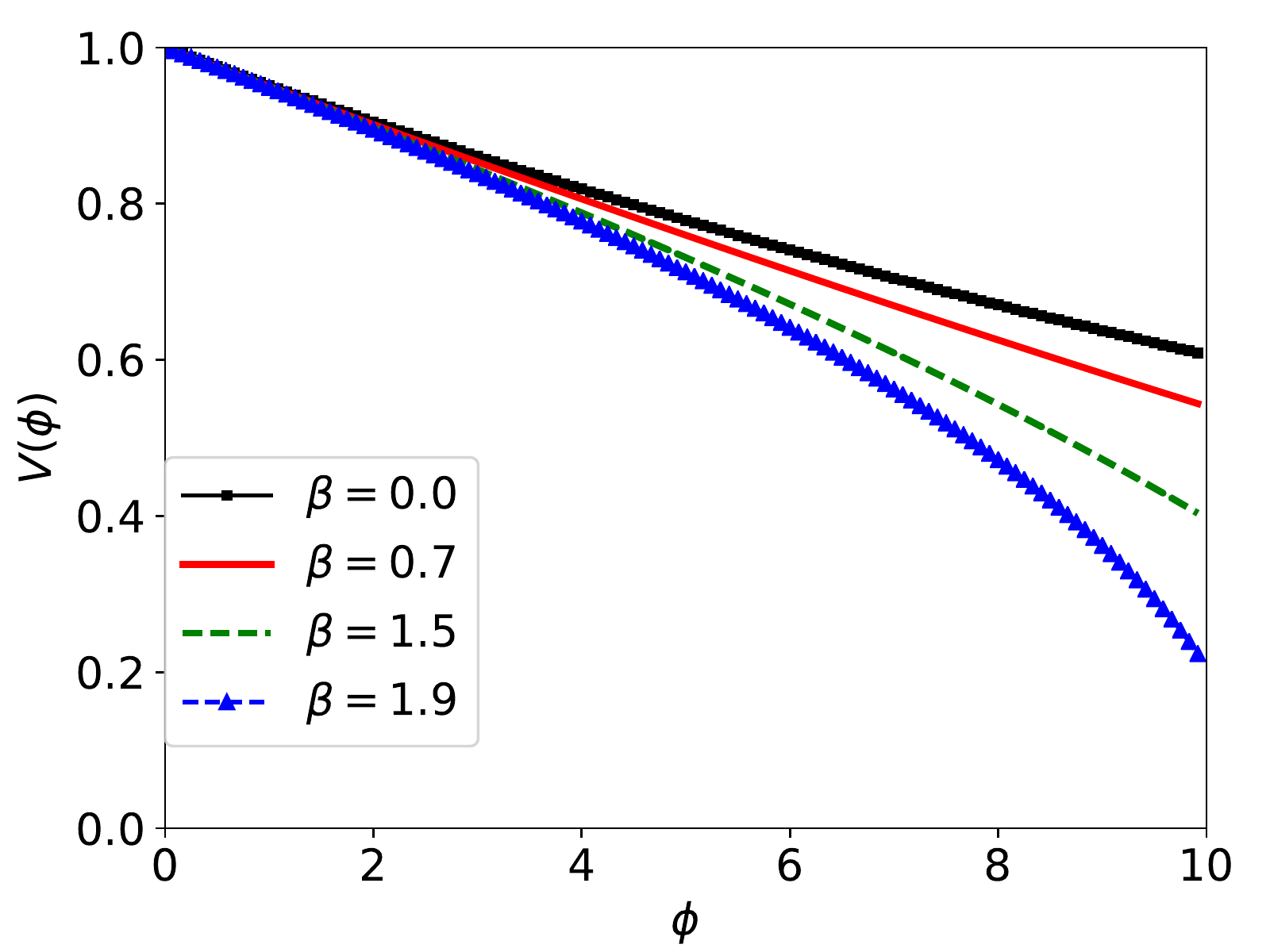}}
\caption{ The potential $V(\phi)$ in function of the field as in Eq. (\ref{potexp}). The value of the parameter $\lambda$ has been fixed at $0.1$ (left panel) and $0.01$ (right panel).}
\label{Fig.1}
\end{figure*}

The brane inflation scalar potential can be readily found from the effective potential into (\ref{action-4d}) and can be written as the $\beta$-exponential potential  of Eq. (1), with $ c_1L=-\beta L/\lambda$,  $L=\lambda^2 \phi$ and $\phi=M_{\rm Pl}^2L$ being the inflaton field. It is interesting to notice that $\beta\in [1/2,\infty)$ --- {{recall that we have previously identified $\beta\equiv\lambda c_1=(\ell^2+1)/2$}, where $b_1=\ell b_2$ and $\ell\in[0,\infty)$}. This agrees with the phenomenologically favored values of the parameter $\beta$ and with the fact that our brane scenario admits  four-dimensional gravity in a limited scale $L\ll r_c$ which is a consequence of the geometry having infinite volume for arbitrary large values of the fifth coordinate $r$. Then the radion  (inflaton) cannot stabilize at infinity as the exponential potential at the limit $\beta\to0$ would require. The finite vacuum expectation values where the radion stabilizes is given by the zeroes of (1), i.e., at $\phi_0=1/\beta\lambda$. For $\lambda=1$ and $\beta=1/2, 1/4$ we find the well-known quadratic and quartic potentials which are usually employed in chaotic inflation. In Fig.~\ref{Fig.1} we show the behaviour of the potential (1) as a function of the field $\phi$ for selected values of the parameters $\beta$ and $\lambda$.
\section{$\beta$-exponential inflation}
\label{Model}

In this section, we discuss some of the theoretical predictions of the class of potential (\ref{potexp}) derived in the previous section. As is well known, the slow-roll inflationary regime is characterized by parameters which depend on the form of the potential and its derivative with respect to the field $\phi$ \cite{book}. For the $\beta$-exponential potential, the slow-roll parameters are written as
\begin{equation} \label{e}
\epsilon(\phi) = \frac{\lambda^2 }{2}\frac{1}{\left[1 - \beta \lambda \phi \right]^2} \quad {\mbox{and}} \quad
\eta(\phi) = \frac{\lambda^2 }{2} \frac{1-2\beta}{\left[1 - \beta \lambda \phi \right]^2}\; .
\end{equation}
In this regime, the end of the inflationary phase is expected to happen at $\phi_e$, where the condition $\epsilon(\phi_e) \sim 1$ is satisfied. From the above equation, we find
\begin{equation}
\phi_e \sim \frac{1}{\beta}\left[\frac{1} {\lambda} - \frac{1}{\sqrt{2}}\right].
\end{equation}
The primordial power spectrum of the curvature perturbation is given by
\begin{equation}
P_R =\frac{V(\phi)}{24\pi^2\epsilon}|_{k=k_{*}}\;,
\label{eq:PR}
\end{equation}
where $(\ast)$ refers to pivot scale, i.e., when the CMB mode exits from horizon at the scale $\phi_{\ast}$. The value of $P_{R}(k_{\ast})$ is set by the COBE normalization to the value $ 2.2 \times 10^{-9}$ for the pivot choice $k_{\ast}=0.05~\Mpc^{-1}$ \cite{Ade:2015xua}. Now, by inverting the Eq.~(\ref{eq:PR}), we can write the value of the amplitude $V_0$  as
\begin{equation}
V_0=\frac{12\pi^2\lambda^2P_{R}(k_{\ast})}{(1-\beta\lambda{\phi_{\ast}})^\frac{1+2\beta}{\beta}}.
\label{eq:V0}
\end{equation}
Noteworthy is the strict dependence of $V_0$ with both the $\lambda$ and $\beta$ parameters, or rather the degeneracy of such parameters in the value of the potential amplitude. This will be of crucial importance for the analysis in the next section. The observable field value $\phi_{\ast}$ can be related to the number of e-folds at which the pivot scale crossed out the Hubble radius during inflation, defined as $N_{\ast}=\int^{\phi_{\ast}}_{\phi_{end}}{d\phi}/{\sqrt{2\epsilon}}$. From the above equations, one finds
\begin{equation}
N_{\ast} = \frac{\beta}{2}\phi_{\ast}^2-\frac{\phi_{\ast}}{\lambda}+ \frac{1}{2\lambda^2\beta}-\frac{1}{4\beta} \;,
\label{eq:efold_b}
\end{equation}
or still,
\begin{equation}
\phi_{\ast}=\frac{1}{\beta\lambda}-\frac{1}{\beta}\sqrt{0.5+2\beta N_{\ast}}\;.
\label{eq:phi_star}
\end{equation}

\begin{figure*}[t]
	\centering{\includegraphics[width=0.45\linewidth]{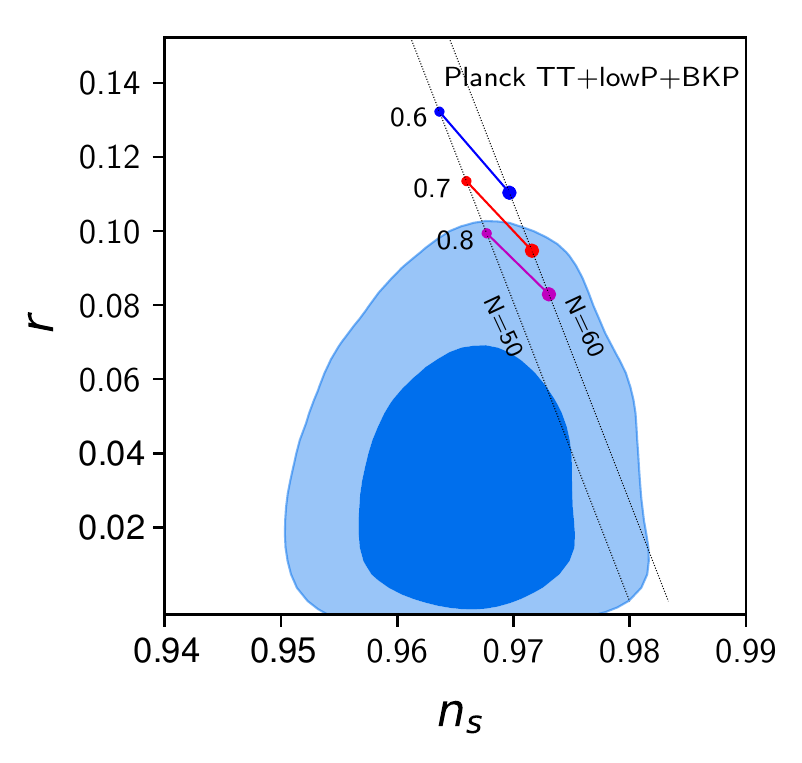}}
	\centering{\includegraphics[width=0.45\linewidth]{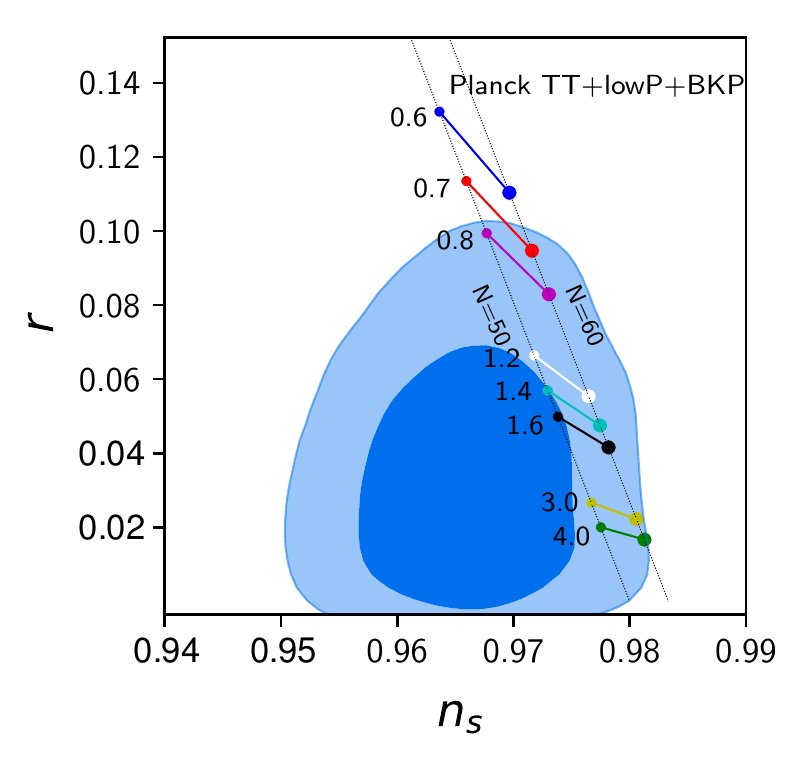}}
	\caption{ The $n_s - r$ plane for the range of values of the parameter $\beta$ satisfying Eq. (\ref{eq:efold_b}), considering two values for the number of {\sl e-folds}, $N = 50$ and $N = 60$. The contours correspond to the Planck(2015)+BICEP2/Keck data ($68\%$ and $95\%$ C.L.) using the pivot $k_{\ast}=0.05~\Mpc^{-1}$. Left panel assumes $\lambda = 0.1$ whereas in the right panel the value of $\lambda$ is fixed to $0.01$.}
	\label{Fig:ns-r}
\end{figure*}

Note that, assuming a positive inflationary field, the slow-roll conditions are fully met by the $\beta$-exponential potential for values of $\beta \geq 0$ and $0 \geq \lambda \geq \sqrt{2}$. Finally, the primordial spectral index, $n_s$, and the tensor-to-scalar ratio, $r$, can be expressed respectively as
%
$n_s-1=-4\epsilon+2\eta$
 and $r=16\epsilon$~\cite{book}.
%
Here, these parameters take the form
\begin{equation} \label{ns}
n_s=1-{\lambda^2}\frac{(1 + 2\beta)}{\left[1 - \beta \lambda \phi_{\ast} \right]^2} \quad {\mbox{and}} \quad
%
r = \frac{8\lambda^2 }{\left[1 - \beta \lambda \phi_{\ast} \right]^2}\;,
\end{equation}
and the relation between $n_s$ and $r$ is given by
\begin{equation} \label{rr}
r = \frac{8(1 - n_s)}{(1+2\beta)}\;.
\end{equation}
Fig. \ref{Fig:ns-r} shows the $n_s-r$ plane for values of $\beta$ satisfying Eq. (\ref{eq:efold_b}) and some selected values of $\lambda$, considering two different numbers of e-folds, i.e., $N = 50$ and $N = 60$. The contours correspond to $68\%$ and $95\%$ (C.L.) obtained from the Planck(2015)+BICEP2/Keck Array data~\cite{Ade:2015lrj}. In an opposite way as the prediction of the spectral index $n_s$, we note that the higher the values of $\beta$ the smaller the model prediction of $r$, which is in agreement with observations. At the same time, the values of the parameter $\beta$ must obey the constraint $\phi_* \geq \phi_{ini}$ for a fixed value of $N_*$, implying in a more restrict interval of $\beta$ (see Eq.~(\ref{eq:phi_star}))\footnote{The initial scale of the slow roll inflation is conventionally assumed to occur at $N_*=70$. Using such a value in Eq.~(\ref{eq:phi_star}) it is possible to calculate $\phi_{ini}$.}. This conclusion changes if lower values of the parameter $\lambda$  is assumed, as can be seen from a direct comparison of the left and right panels of Fig.~(\ref{Fig:ns-r}). 
{{Worthing note that the $\beta$-inflation predictions allow for tensor-to-scalar ratio consistent with Planck results at the 1-$\sigma$ C.L. for $\beta \gtrapprox 1.2$, while the spectral index is never compatible with 1-$\sigma$ C.L. for such values of $\beta$ parameter. These predictions are very close to others already studied in literature, such as those of single monomial potentials \cite{Ade:2015lrj}.}}

\begin{table}[t]
	\centering
	\caption{ Priors on the cosmological parameters considered in the analysis.}
	\vspace{0.2cm}
	\begin{tabular}{|c|c|}
		\hline
		Parameter  & Prior   \\
		\hline
		
		$100\,\Omega_b h^2$ & $(0.005,  0.1)$   \\
		
		$\Omega_{c} h^2$ & $(0.001,  0.99)$    \\

		$100\, \theta$ & $(0.5,  10)$ \\
		
		$\tau$    & $(0.01, 0.8)$ \\
		
		$\beta$    & $(0.3, 5)$ \\
		
		$\lambda$\footnote{Prior for the $\beta$-inflation model-1.}    & $(0.01, 0.17)$ \\
		\hline
	\end{tabular}
	\label{prior}
\end{table}
%
\section{Method and Analysis}
\label{Method}
We perform a Bayesian model comparison analysis considering three models, namely, the standard $\Lambda$CDM scenario (as reference model) and the $\beta$-exponential inflation model varying both the $\beta$ and $\lambda$ parameters and fixing $\lambda$ 
to a proper value, hereafter ``model-1" and ``model-2", respectively.

\begin{table*}[t]
\centering
\caption{$68\%$ confidence limits for the cosmological parameters using the TT+lowP Planck (2015) data.
The first column shows the constrains on the reference ${\Lambda}$CDM model whereas the second and third columns show, respectively, the results of the  analysis for the $\beta$-inflation model varying both the $\beta$ and $\lambda$ parameters and for the $\beta$-inflation model with $\lambda$ fixed at the arbitrary value of $0.07$. The $\Delta \chi^2_{best}$ and the $\ln \mathcal{B}_{ij}$ refer to the difference between the model and the $\Lambda$CDM analysis.}
\label{tab:Tabel}
\begin{tabular}{|c|c|c|c|}
\hline
Parameter &${\Lambda}$CDM & $\beta$-infl. (model-1) & $\beta$-infl. (model-2)\\
\hline
\hline
$100\,\Omega_b h^2$ 	
& $2.222 \pm 0.022$  		
& $2.245 \pm 0.019$  		
& $2.247 \pm 0.019$			
\\
$\Omega_{c} h^2$	
& $0.1197 \pm 0.0021$ 	
& $0.1167 \pm 0.0012$ 
& $0.1163 \pm 0.0012$
\\
$100\, \theta$
& $1.04085 \pm 0.00045$ 	
& $1.04120 \pm 0.00041$ 
& $1.04130 \pm 0.00041$
\\
$\tau$
& $0.077 \pm 0.018$	
& $0.094 \pm 0.016$	
& $0.097 \pm 0.017$
\\
$n_s$
& $0.9655 \pm 0.0062$ 	
& $ - $ 
& $ - $ 	
\\
$\ln ( 10^{10}A_s )$  
& $3.088 \pm 0.034$
& $ - $ 
& $ - $
\\
$\beta$
& $ - $ 				
& $ 1.63 \pm 0.57 $ 				
& $ 1.92 \pm 0.05$		
\\
$\lambda$
& $ - $ 				    
& $ 0.079 \pm 0.013$ 	
& fixed to $0.07$		
\\
\hline
\hline
$H_0 $ {[km s$^{-1}$ Mpc$^{-1}$]}
& $67.31 \pm 0.95$
& $68.68 \pm 0.54$ 
& $68.86 \pm 0.52$
\\
$\Omega_m$ 	
& $ 0.315 \pm 0.013$  		
& $ 0.296 \pm 0.007$  		
& $ 0.294 \pm 0.007$ %
\\
$\Omega_{\Lambda}$ 	
& $ 0.685 \pm 0.013$  		
& $ 0.703 \pm 0.007$  		
& $ 0.706 \pm 0.007$	
\\
\hline
\hline
$\chi^2_{best}$
& $11263.0$ 
& $11266.9$ 
& $11266.2$
\\						
$\Delta\chi^2_{best}$
& $-$
& $ - 3.9 $ 
& $ - 3.2 $  
\\
\hline
\hline
$\ln \mathcal{B}$
& $-5674.0$
& $-5682.3$
& $-5676.4$
\\
$\ln \mathcal{B}_{ij}$ 
& $-$
& $ - 8.3$
& $ - 2.4 $
\\
\hline
\end{tabular}
\end{table*}
In order to perform our analysis, we use the {\sc CosmoMC} code~\cite{Lewis:2002ah} and the {\sc MultiNest} algorithm~\cite{Feroz:2008xx,Feroz:2007kg,Feroz:2013hea}, necessary to resolve the Boltzmann equations, explore the cosmological parameter space and make a Bayesian models selection. Two main modifications to the most recent {\sc CosmoMC} release are performed. The first is in the Code for Anisotropies in the Microwave Background ({\sc CAMB})~\cite{camb}, already included in the {\sc CosmoMC}, since in its basic realization it assumes a power-law parametrization for the primordial perturbation spectrum as $P_R=A_s (k/k_*)^{ns-1}$.
Instead, in this work we intend to use the $\beta$-inflationary primordial potential form given by Eq.~ (\ref{potexp}),
and we need to compute the dynamics and perturbations of this model to construct the primordial power spectrum. In this context, we propose a modification in {\sc CAMB} following the lines of the~{\sc ModeCode}~\cite{Mortonson:2010er, Easther:2011yq} adapted for the our primordial potential choice. This latter code is able to compute the CMB anisotropies spectrum solving numerically the inflationary mode equations, i.e., solving the Friedmann and Klein-Gordon equations as well as the
Fourier components of the gauge-invariant quantity $u$ for an exact form of the single field inflaton potential $V(\phi)$.
By integrating these equations it is possible to obtain $H$ and $\phi$ as a function of time and the solution $u_k$ for the mode $k$. Therefore, following these steps, the code can compute the power spectrum of the curvature perturbation $P_\mathcal{R}$ by $P_\mathcal{R} = \frac{k^3}{2\pi^2}\left|\frac{u_k}{z}\right|^2$, evaluated when the mode crosses the horizon.

The second main modification is made in the {\sc CosmoMC} source, i.e., by implementing the nested sampling of the code {\sc MultiNest}~\cite{Feroz:2008xx,Feroz:2007kg,Feroz:2013hea} to achieve our Bayesian analysis of the model.
The code {\sc MultiNest} is able to accurately analyze models with high number of parameters and non-gaussian density distributions and/or pronounced degeneracies. It also calculates the Bayesian evidence of the model, allowing the Bayesian model comparison. In this model selection, the ``best'' model is the one that achieves the best compromise between quality of fit of the data and predictivity, which means that the model that better fits the data thanks to many free parameters must be weighed with its added complexity  (we refer the reader to \cite{Liddle:2007fy,Trotta:2005ar, Santos:2016sti, Campista:2017ovq,Benetti:2016tvm,SantosdaCosta:2017ctv, Graef:2017cfy, Benetti:2016ycg} for some recent applications of Bayesian model selection in cosmology).
In our analysis, we vary the usual cosmological parameters, namely, the physical baryon density, $\Omega_bh^2$, the physical cold dark matter density, $\Omega_ch^2$, the ratio between the sound horizon and the angular diameter distance at decoupling, $\theta$, the optical depth, $\tau$ and the parameters $\beta$ and $\lambda$.
We also vary the nuisance foreground parameters~\cite{Aghanim:2015xee} and consider purely adiabatic initial conditions. The sum of neutrino masses is fixed to $0.06$ eV, and we limit the analysis to scalar perturbations with $k_*=0.05$ $\rm{Mpc}^{-1}$. We perform our analysis assuming the priors on the cosmological parameters shown in Tab.~\ref{prior}. The values of the parameters $\beta$ and $\lambda$ are chosen from the considerations made in the previous section --  in particular we refer to the observational predictions of Fig.~\ref{Fig:ns-r} and also recall the correlation between the parameters $\lambda$ and $\beta$, as shown in Eq.~(\ref{eq:V0}). Mainly for this reason, we perform two kind of analysis, leaving both the parameters free to vary (model-1) and fixing the $\lambda$ value to a proper value (model-2).
In particular, the $\lambda$ value we consider for the model-2 is the best fit obtained of such parameter in the model-1 analysis.
For both analyzes, it is assumed the arbitrary value of the number of e-folds, i.e., $N_{\ast}=55$.
\begin{table*}[!]
	\centering
	\caption{$68\%$ confidence limits for the cosmological parameters using the TT+lowP+HST. The first column shows the constrains on the reference ${\Lambda}$CDM model whereas the second shows the results of the analysis on the $\beta$-inflation model with $\lambda$ fixed to the arbitrary value of $0.07$. As in Table II, the $\Delta \chi^2_{best}$ and the $\ln \mathcal{B}_{ij}$ refer to the difference between the model and the $\Lambda$CDM analysis.}
	\label{tab:Tabe2}
		\begin{tabular}{|c|c|c|}
			\hline
			Parameter &${\Lambda}$CDM (TT+lowP+HST) & $\beta$-infl. model-2 (TT+lowP+HST)\\
			\hline
			\hline
			$100\,\Omega_b h^2$ 	
			& $2.245 \pm 0.022$  		
			& $2.253 \pm 0.019$			
			\\
			$\Omega_{c} h^2$	
			& $0.1167 \pm 0.0019$ 	    
			& $0.1157 \pm 0.0011$       
			\\
			$100\, \theta$
			& $1.04130 \pm 0.00044$ 	
			& $1.04141 \pm 0.00040$     
			\\
			$\tau$
			& $0.091 \pm 0.019$	        
			& $0.098 \pm 0.016$         
			\\
			$n_s$
			& $0.9730 \pm 0.0057$ 	
			& $ - $ 	
			\\
			$\ln ( 10^{10}A_s )$  
			& $3.109 \pm 0.036$
			& $ - $
			\\
			$\beta$
			& $ - $ 				
			& $ 1.92 \pm 0.05$		
			\\
			$\lambda$
			& $ - $ 				    
			& fixed to $0.07$		
			\\
			\hline
			\hline
			$H_0 $ [km s$^{-1}$ Mpc$^{-1}$]
			& $68.74 \pm 0.87$
			& $69.19 \pm 0.49$
			\\
			$\Omega_m$ 	
			& $ 0.296 \pm 0.011$  		
			& $ 0.290 \pm 0.006$ %
			\\
			$\Omega_{\Lambda}$ 	
			& $ 0.704 \pm 0.011$  		
			& $ 0.710 \pm 0.006$	
			\\
			\hline
			\hline
			$\chi^2_{best}$
			& $11272.9$ 
			& $11271.0$ 
			\\						
			$\Delta\chi^2_{best}$
			& $-$
			& $1.9 $  
			\\
			\hline
			\hline
			$\ln \mathcal{B}$
			& $-5682.6$
			& $-5680.0$
			\\
			$\ln \mathcal{B}_{ij}$
			& $-$
			& $ 2.6 $
			\\
			\hline
	\end{tabular}
\end{table*}
We use the second release of Planck data \cite{Aghanim:2015xee} (hereafter TT+lowP), namely, the high-$\ell$ Planck temperature data (for $30< \ell <2508$) from the 100-,143-, and 217-GHz half-mission TT cross-spectra and the low-P data by the joint TT, EE, BB and TE likelihood (in the range of $2< \ell <29$). For the second analysis, we also consider the Riess {\it{et al.}} results on the local expansion rate, $H_0 = 73.24 \pm 1.74$ $\rm{km.s^{-1}.Mpc^{-1}}$ (68\% C.L.), based on direct measurements made with the Hubble Space Telescope \cite{Riess:2016jrr}. This measurement is used as an external Gaussian prior and we refer to this joint data set as TT+lowP+HST.

It is worth mentioning that for our results we use the most accurate Bayesian Importance Nested Sampling (INS)~\cite{Cameron:2013sm, Feroz:2013hea} instead of the vanilla Nested Sampling (NS), requiring INS Global Log-Evidence error $< 0.1$. In order to rank the models of interest, we use the scale in terms of the evidence strength of the chosen reference model~\cite{Trotta:2005ar}: $ \ln{\mathcal{B}_{ij}} = 0 - 1 $ , $\ln{\mathcal{B}_{ij}}  =1 - 2.5$, $ \ln{\mathcal{B}_{ij}}  =2.5 - 5$, and $ \ln{\mathcal{B}_{ij}}  >5 $ indicate, respectively, an {\textit{inconclusive}}, {\textit{weak}}, {\textit{moderate}} and {\textit{strong}} preference of the model $i$ with respect to the reference model $j$. Note that negative values of $\ln{\mathcal{B}_{ij}}$ mean support in favour of the reference model.

\begin{figure*}[t]
	\centering
	\includegraphics[width=0.65\hsize]{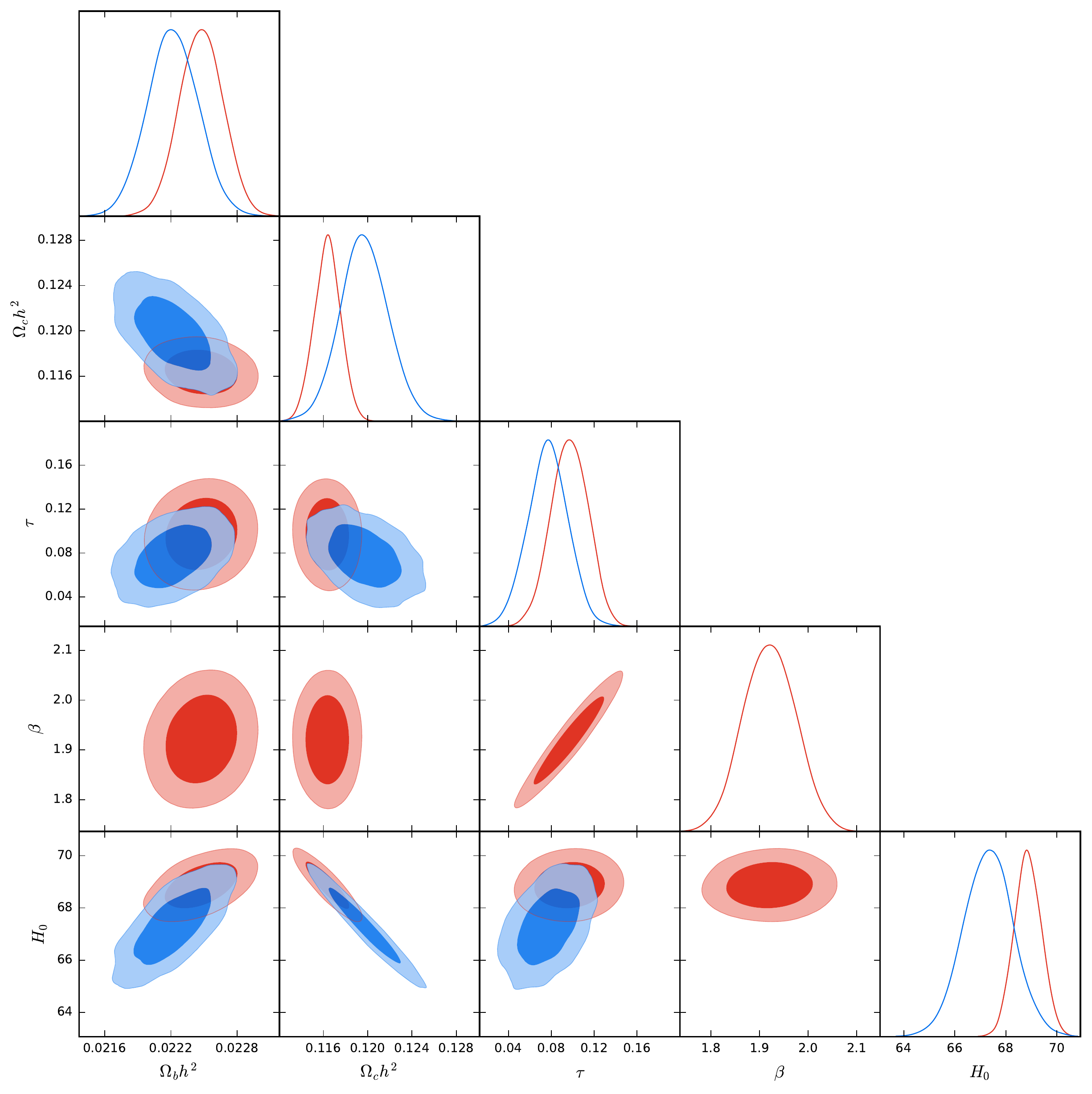}
	\caption{Confidence regions for the $\beta$-inflation `model-2' analysis (red contours) and the reference $\Lambda$CDM model (blue contours), both using the TT+lowP Planck (2015) data.}
	\label{fig:triangle_TT}
\end{figure*}
%
\section{Results}
\label{Results}

\begin{figure*}[t]
\centering
\includegraphics[width=0.7\hsize]{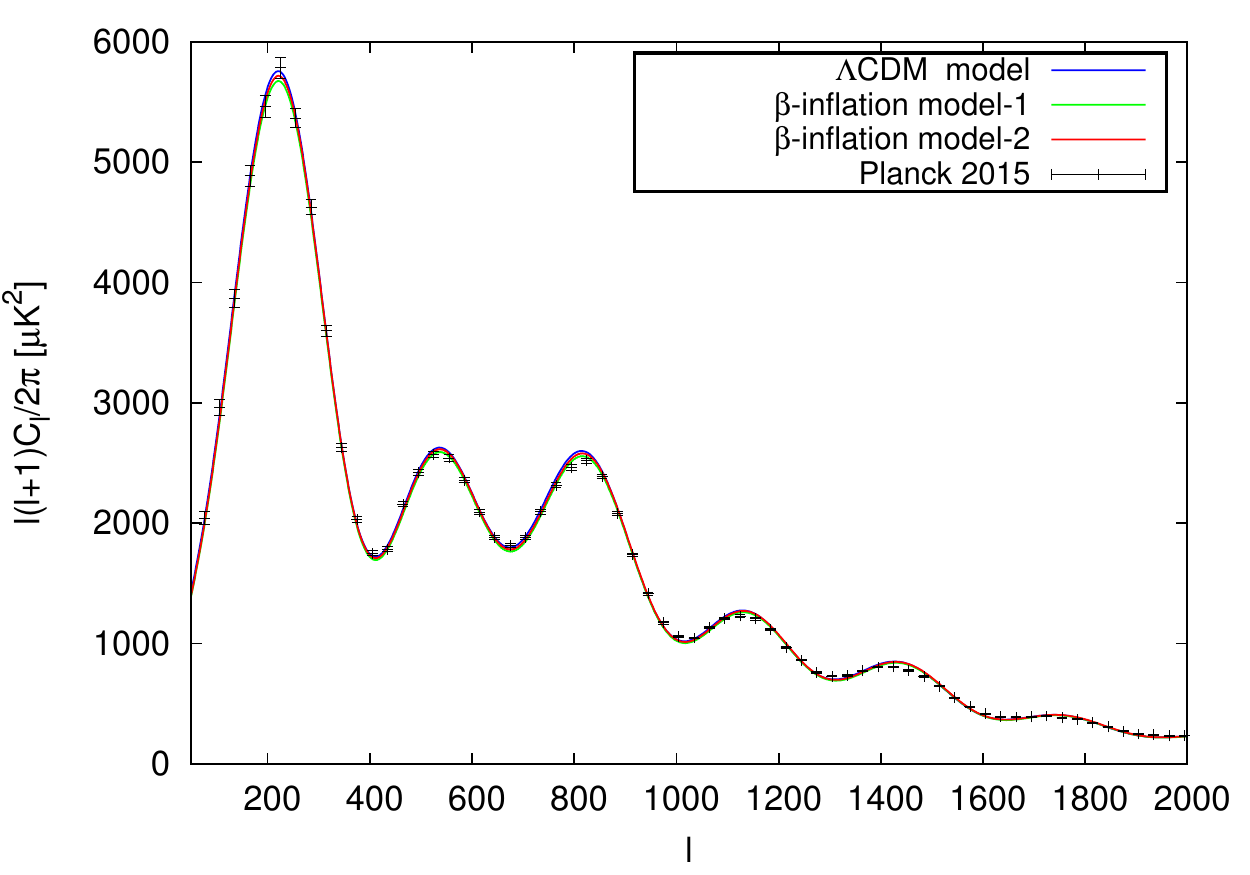}
\caption{Temperature power spectrum for the $\beta$-inflation model-2 best fit values (red curve)and  $\beta$-inflation model-1 (green curve) in comparison with the $\Lambda$CDM model best fit (blue curve) and the Planck (2015) data.}
\label{fig:bestfit_TT}
\end{figure*}

The main quantitative results of our analysis using the TT+lowP data are shown in Tab.~\ref{tab:Tabel}, where we report the constraints on the cosmological and primordial parameters for the three analyzed scenarios. Note that the results for the $\beta$-inflation models agree with the $\Lambda$CDM predictions at $1\sigma$, as also shown in Fig.~\ref{fig:triangle_TT}. In Fig.~\ref{fig:bestfit_TT} we plot the best fit curves for the analyzed models in comparison with the $\Lambda$CDM cosmology. In the last line of Tab.~\ref{tab:Tabel} we can see that the $\beta$-inflation model-1 is \textit{strongly} disfavoured with respect to the reference model. At the same time, in Tab.~\ref{tab:Tabe2} we report the results using the TT+lowP+HST data. We can see that the choice of fixing the $\lambda$ value (model-2) allows to better constrain  the $\beta$ value, decreasing the model comparison to \textit{weakly} disfavoured. This is basically due to the parameter degeneracy between $\beta$ and $\lambda$, discussed in the previous sections.

An important aspect that is worth mentioning concerns the predictions on the spectral index, $n_s$, of $\beta$-inflation model (see Fig.~\ref{Fig:ns-r}). This class of models predicts a higher value of $n_s$ with respect to the $\Lambda$CDM model for the $\beta$ values constrained by the data. In particular, the derived value of the spectral index is $n_s \sim 0.976$ for the model-1 and $n_s \sim 0.977$  for the model-2 (see Eq.~\ref{ns}). As is well known (for a comprehensive reading, see Ref.~\cite{Benetti:2017gvm}), the higher the value of $n_s$ the higher the values of $\Omega_bh^2$ and $\tau$ and, consequently, the higher value of $H_0$. Therefore, this inflationary model naturally leads to higher values of $H_0$, which is in better agreement with the local expansion rate based on direct measurements made with the Hubble Space Telescope \cite{Riess:2016jrr}.

As mentioned earlier, we also perform an analysis of the $\beta$-inflation model in light of the joint  TT+lowP+HST data set. The results are shown in Tab.~\ref{tab:Tabe2}. We note that the prior on $H_0$ modifies the previous results for the evidence of the standard model and makes the $\beta$-inflation $moderately$ preferred ($\ln{\mathcal{B}_{ij}} = 2.6$) over the reference $\Lambda$CDM cosmology. 
It is important to emphasize that these results depend on the choice of the e-fold number used for the analysis, i.e, $N_{\ast}=55$. For a lower value of $N_{\ast}$, one may obtain a better agreement with the $\Lambda$CDM model, which would be in accordance with the results shown in Fig.~(\ref{Fig:ns-r}).
{Remarkable, from a results comparison of the analysis using only CMB data, Tab. \ref{tab:Tabel}, and CMB+HST data, Tab. \ref{tab:Tabe2}, we note that adding the $H_0$ prior makes the Bayesian evidence and absolute log likelihoods worst. In this context, the $\beta$ inflation model partially compensates this by describing the data better than the $\Lambda$CDM model does.}


\section{Conclusions}
\label{Conclusions}

Cosmic Microwave Background data are one of the most powerful tools to study the early universe physics. In particular, the past two decades have witnessed a great improvement in the measurements of the CMB fluctuations, which are now able to test the observational viability or even rule out different classes of inflationary models  as well as some their alternatives.

In this paper, we have analyzed theoretical and observational aspects of a particular class of inflationary models proposed in Ref.~\cite{alcaniz}, whose field potential is given by Eq.~(\ref{potexp}). First, we have shown that this kind of potential can arise in the context of brane cosmology, where the radion is interpreted as the inflaton field. The observational viability of this class of models have been studied through a Bayesian analysis using the latest Planck (2015) data.  As shown in Tables II and III, tight constraints on the parameter $\beta$ have been derived. Apart from the value of the spectral index $n_s$, our analysis shows that the predictions of the $\beta$-inflation model are very similiar to the ones of the $\Lambda$CDM model (they agree at 68.3\% C.L.). Considering only the TT+lowP CMB data, we have shown that the minimal standard model is \textit{weakly} preferred over the $\beta$-exponential inflation. However, this result changes when we also consider the most recent HST measurements of the Hubble parameter, as reported in Ref.~\cite{Riess:2016jrr}. In this case (TT+lowP+HST data), the $\beta$-inflation model becomes {\it moderately} preferred over the $\Lambda$CDM cosmology, with $\ln{\mathcal{B}_{ij}} = 2.6$.

Finally, it is important to mention that stringy and brane inspired potentials have been assumed to be the ultimate potentials to accomplish unique inflaton potentials from fundamental theories. But one usually faces problems with the comparison of their theoretical predictions and observational data. In the present study, we have shown that the $\beta$-exponential potential shows to be in agreement with the current observational data at the same time that it can be derived from a fundamental theory such as supergravity with dilatonic braneworld solutions. Another important aspect that is worth emphasizing in the present scenario is the fact that large values of $\beta$ are in agreement with both observational data and radion stabilization at finite size in brane inflation scenario. This somewhat appears as a completion of other recent fundamental potential scenarios discussed in \cite{kallosh}.

\section{Acknowledgments}

M. S. and R. S. acknowledges financial support from Conselho Nacional de Desenvolvimento Cientifico e Tecnol\'ogico (CNPq). M. B. is supported by Funda\c{c}\~ao Carlos Chagas Filho de Amparo \`a Pesquisa do Estado do Rio de Janeiro (FAPERJ) through the post-doc Nota 10 fellowship. JSA acknowledges support from CNPq (Grants no. 310790/2014-0 and 400471/2014-0) and FAPERJ (grant no. 204282).  FAB acknowledges support from CNPq (Grant no. 309258/2014-6). The authors thank the use of COSMOMC and the MULTINEST codes.

\end{document}